\documentclass[showpacs,preprintnumbers,amsmath,amssymb,twocolumn]{revtex4}
\usepackage{mhchem}
\usepackage{graphicx}% Include figure files
\usepackage{dcolumn}% Align table columns on decimal point
\usepackage{bm}% bold math
\usepackage{xcolor}
\def\bea {\begin{eqnarray}}
\def\eea {\end{eqnarray}}

\def\be {\begin{equation}}
\def\ee {\end{equation}}

\begin{document}
\sloppy
\title{Glauber model for small system using anisotropic and inhomogeneous density profile of proton}
\author{Suman Deb}
\author{Golam Sarwar}
\author{Dhananjaya Thakur}
\author{Pavish S.}
\author{Raghunath Sahoo}
\email{Raghunath.Sahoo@cern.ch}
%\medskip
\affiliation{Discipline of Physics, School of Basic Science, Indian Institute of Technology Indore, Simrol, Indore 453552, India}
\author{Jan-e Alam}
\email{jane@vecc.gov.in}
\affiliation{Variable Energy Cyclotron Centre, 1/AF, Bidhan Nagar, Kolkata - 700064, India}

\begin{abstract} 
Recent studies reveal that at high energies, collisions of small system like $p+p$ gives signatures similar to that widely observed in heavy ion collisions hinting towards a possibility of forming a medium with collective behaviour. With this motivation, we have used the Glauber model, which is traditionally applied to heavy ion collisions, in small system using anisotropic and inhomogeneous density profile of proton and found that the proposed model reproduces the charged particle multiplicity distribution of $p+p$ collisions at LHC energies very well. Collision geometric properties like mean impact parameter, mean number of binary collisions ($\langle N_{coll} \rangle$) and mean number of participants ($\langle N_{part} \rangle$) at different multiplicities are determined. Having estimated $\langle N_{coll} \rangle$, we have calculated nuclear modification-like factor ($R_{HL}$) in $p+p$ collisions. We also estimated eccentricity and elliptic flow as a function of charged particle multiplicity using the linear response to initial geometry. 

\end{abstract}
 
\date{\today}
\maketitle
\section{Introduction}

%Relevance of pp collisions
Results of relativistic proton proton ($p+p$) collisions are used as reference or base line for interpreting various results of heavy ion collisions at relativistic energies, which are aimed at creation and characterization of phases of strongly interacting matter governed by Quantum Chromodynamics (QCD). At high temperature/density, a deconfined thermalised state of quarks and gluons called Quark-gluon plasma (QGP) has been predicated by lattice QCD based calculations~\cite{Bazavov:2014pvz,Borsanyi:2013bia}.  Values of ratio  of certain observable measured in heavy ion collisions,  such as number of produced strange particles, production of  $J/\psi$, to those measured in $p+p$ collision are interpreted as  signature of partonic medium formation in heavy ion collisions; e.g., enhancement of number of strange particles and suppression in number of $J/\psi$ in collisions of heavy ion with respect to that of $p+p$ (approximately scales by binary collisions) are taken as signatures of QGP formation in relativistic heavy collisions~\cite{Krzewicki:2011ee,Aamodt:2010cz,Hirano:2010je,Arsene:2004fa,Back:2004je,Adams:2005dq,Shuryak:2004cy,Gyulassy:2004zy,Muller:2006ee}. Apart from taking values of such ratio as  confirmation for creation of QGP in such collisions, they are also used in characterizing QGP as well as in verifying and  constraining different theoretical models. For such interpretations, it is assumed that, in $p+p$ collisions, no partonic medium is formed. However, recent results show that such assumptions may not be correct for high multiplicity $p+p$ collisions~\cite{ALICE:2017jyt,Alver:2010ck,Khachatryan:2010gv}. Understanding of $p+p$ collisions is crucial for characterization of the QCD medium formed in heavy ion collision.
% Aso the fact that value of nuclear suppression factor for heavy ion collision is less than unity is taken as sign of partonic medium. This suppression factor is also defined as ratio of number of produced particles in heavy ion collisions  to that of p+p collisions.
     \begin{figure}
  \includegraphics[scale=0.581]{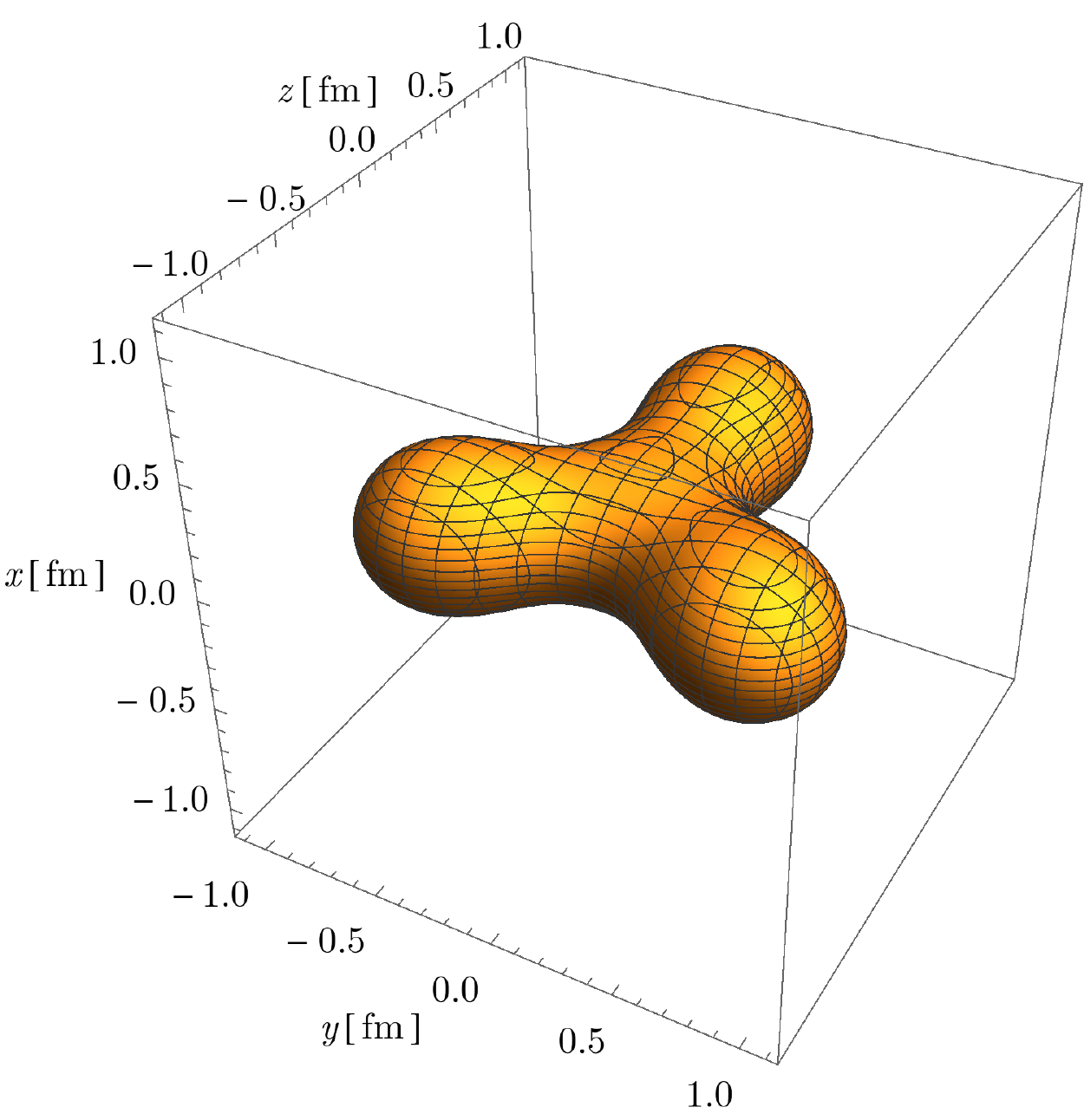}
\caption{(Color online) Depiction of effective quarks and gluonic flux tubes connecting them within a proton~\cite{Kubiczek:2015zha}.}
\label{figA}     
\end{figure}

% Relavance of initial condion
The evolution of matter formed in relativistic heavy ion collisions is critically dependent on the initial conditions. The extraction of the signals of the QGP strongly dependent on the initial conditions~\cite{Yan:2017bgc,Yagi:2005yb,Wong:1995jf}. For example, the elliptic flow ($v_{2}$) of hadrons are calculated using the transverse momentum ($p_{T}$) spectrum of hadrons. The $p_{T}$ spectra are estimated using the hydrodynamical models. To solve the hydrodynamical equations, initial conditions and equation of state are required as inputs. Therefore, the shear viscosity of the system extracted through $v_{2}$ will be sensitive to initial conditions, i.e., any uncertainties in initial condition will be reflected in the extracted value of shear viscosity~\cite{Song:2010mg}.

In high energy heavy-ion collisions, interpretation of results relies on the use of a model based on initial matter distribution resulting from the overlap of the two colliding nuclei at a given impact parameter (b). Indeed, for estimating quantities such as: (i) the centrality dependence of any observable expressed by the number of participating nucleons in the collision, $N_{part}$(b), (ii) the number of binary nucleon-nucleon collisions, $N_{coll}$(b) used to derive the nuclear modification factor ($R_{AA}$) from the ratio of AA over pp spectra, (iii) the elliptic and triangular flow parameters ($v_2$) and ($v_3$) normalized by the eccentricity $\epsilon_2$(b) and triangulation $\epsilon_3$(b) of the overlap region, and (iv) the average surface area, A(b) and (v) path length, L(b) of the interaction region, knowing the nuclear overlap function $T_{AA}$(b) is important. And, this overlap function depends on a realistic model of the collision geometry~\cite{Miller:2007ri}.
 
Similar to heavy ion collisions, it is imperative to understand the initial condition of the medium formed in $p+p$ collisions for high multiplicity events. Apart from this, knowing proper initial condition can also give a possible way to define centrality classes and the base needed for properly defining suppression factors or ratios for comparing results of event of different multiplicity classes produced in p+p collisions~\cite{Adam:2014qja}. Appropriate initial conditions can be chosen by considering  that it should reproduce certain aspects of results such as multiplicity distribution or centrality distribution of various observable related to the events. 
 
% Where to get what are the possibilities
For constructing proper initial conditions for p+p collisions, at first attempt, one follows the way similar to that of heavy ion collisions. Initial conditions for heavy ion collisions are modeled in two kinds of distinct approaches: (i) one considers nucleonic or partonic collisions for energy deposition in the collision zone, those are based on Glauber model~\cite{Kubiczek:2015zha}, and, (ii) QCD based calculations are employed to estimate initial energy deposition by gluonic fields originated from partonic currents of colliding nuclei \cite{Schenke:2012wb}. So these will also be obvious approaches for modeling initial conditions in p+p collisions. As models based on Glauber modeling, are very successful in reproducing various results of relativistic heavy ion collisions, one can consider models for initial conditions of $p+p$ collisions which are based on similar kind of assumptions as used for Glauber approach used in heavy ion collisions.

%Where are the problems, what can be improved 
% Which things are required
% What has been done 

Initial transverse shape of the nuclei as described by Glauber model for heavy-ion collisions depends on Wood-Saxon distribution, which is a two-parameter (half-density radius (R) and diffusivity (a)) Fermi-like distributions (2pF) extracted from fits to elastic lepton-nucleus data~\cite{DeJager:1974liz,DeJager:1987qc}, which describes the multi-nucleon interactions occurring in the overlap region between the colliding nuclei via a Glauber eikonal approach~\cite{Glauber:1970jm}. Whereas, in the Monte Carlo Glauber (MCG) models~\cite{Wang:1991hta,Alver:2008aq,Loizides:2014vua,Broniowski:2007nz,Rybczynski:2013yba,Loizides:2016djv}, event-by-event sampling of individual nucleons are done from Wood-Saxon distribution and averaging over multiple events are used to calculate properties related to collisions. Presently, available partonic Glauber model for p+p collisions does not consider full anisotropic density  profile of protons, though radial homogeneity is assumed.

%In the Monte Carlo Glauber (MCG) models (e.g., Refs. [Ref]),individual nucleons are sampled event-by-event from the underlying 2pF distributions and the collision properties are calculated by averaging over multiple events. However, neutron-rich nuclei such as \ce{^{208}Pb} may have differing proton and neutron density distributions at the nuclear periphery.Indeed, measurements have recently been able to extract the neutron profile of several nuclei that show differences with respect to their proton distribution [Ref], and various works have already studied its impact on different isospin-dependent observables in nuclear collisions [Ref]. Presently, available partonic Glauber model for p+p collisions does not consider full anisotropic density  profile of protons, though radial homogeneity is assumed.

In this article, we present the results of  Glauber-like model calculations for $N_{coll}$(b), $N_{part}$(b) due to the quark and gluon based proton density profile, which is a realistic picture obtained by results of deep inelastic scattering that reveals the structure of proton~\cite{Kubiczek:2015zha}, and used it to obtain charged particle multiplicity distribution in $p+p$ collisions at $\sqrt{s}$ = 7 TeV.  Calculated multiplicity distribution is contrasted with ALICE data, a relation of impact parameter with multiplicity is calculated and multiplicity distribution of eccentricity and flow harmonics is estimated for $p+p$ collisions. In order to understand the possibility of medium formation in high-multiplicity $p+p$ collisions, we have estimated nuclear modification-like factor, $R_{HL}$, considering low multiplicity yields as the base.
%Results are also compared with other Glauber models for p+p collisions with different kind of proton density profile. [FUTURE WORK]   

%In the present work we make an attempt to understand
%the evolution of the QGP created in HIC through the study of fluctuations
%in variaous thermodynamical quantities within the framework of Boltzmann 
%equation~\cite{degroot}.

The paper is organized as follows. In the next section we  discuss the formalism that is used in this work. In section III, we present the results and section IV is devoted to summary and discussions.

%===============================================================================
\section{GLAUBER FORMALISM}  
%The phase space distribution function
%---------------------------------------------
In the literatures, the density profiles like hard sphere, and 2pF functions are used traditionally to formulate Glauber model for heavy ion and even for protons~\cite{Loizides:2016djv}. All these profiles can also be extended to proton model by considering radially symmetric parton density. In fact, in the case of proton, several density profiles have been considered to estimate the initial conditions, most of them assume azimuthally symmetric density profile, those are mainly different in phenomenological parameterization of radial variations~\cite{dEnterria:2010xip}. But the standard model postulates that a proton consists of three effective quarks (constituent) and gluons within it. Thus distribution of such configuration is less likely to be radially symmetric, because we expect individual peaks in wave function in the quarks position inside a proton indicating its presence. The necessary condition is, however, that the wave function of each effective quarks and gluons should decay rapidly around boundary of a proton (within RMS area).  In this regard we find only one previous work~\cite{Kubiczek:2015zha} to consider azimuthally asymmetric and inhomogeneous density distribution of proton ~\cite{Bjorken:2013boa, Glazek:2011wf}, which is motivated by  the shape of structure function obtained in deep inelastic scattering, pointing out that multiplicity distribution produced by different model can be used to discriminate them-which can better reproduce experimental results. 
 The difference between the present work and that of reported in Ref. ~\cite{Kubiczek:2015zha} are: (i)	We have considered different possible configurations of Gaussian-fluctuating model thereby ensuring each collision as unique by assuming initial position vectors of the three quarks to be vertices of an equilateral triangle in xy-plane and then in order to account for all possible configurations, position vectors of the quarks are parameterized by varying azimuthal and polar angles. This parameterization is done by considering tilts of the quarks' initial configuration by some angle along x-axis followed by rotation of some other angles. Similar approach is applied along y-axis as well. In this process of parameterization, angles chosen in such a way, that there is no repetition of the particular configuration,
(ii) For the estimation of charged particle multiplicity ($N_{ch}$), in Ref.~\cite{Kubiczek:2015zha}, it is assumed that $N_{ch}$ for each event is in linear scaling with number of binary collisions ($N_{coll}$). But, in the present work, we have considered contribution of number of participants ($N_{part}$) along with ($N_{coll}$) for the estimation of charged particle multiplicity as $N_{part}$ dominates low-$p_T$ region and $N_{coll}$ contribution is higher in the high-$p_T$ domain. A combination of both, which is our approach, appears to be more reasonable. We have calculated elliptic flow using linear response to eccentricity.

In this study, we have used a model with fluctuating proton orientation and it has three effective quarks and gluonic flux tubes connecting them as shown in Fig.~\ref{figA}. The densities of quarks ($\rho_{q}$) and gluons ($\rho_{g}$) are taken as Gaussian type assuming spherically symmetric distribution of quark densities from their respective centers and cylindrically symmetric gluon densities about the line joining two adjacent quarks as 

\begin{equation}
\rho_{q} (\textbf{r} ; r_{q}) = \frac{1}{(2\pi)^{3/2}r_{q}^3} e^{-\frac{r^2}{2r_{q}^2}}
\label{rho_quark}
\end{equation} 

\begin{equation}
\rho_{g} (\textbf{r} ; r_{s} , r_{l}) = \frac{1}{(2\pi)^{3/2}r_{s}^2r_{l}} e^{-\frac{x^2 + y^2}{2r_{s}^2} - \frac{z^2}{2r_{l}^2}}
%\end{split}
\label{rho_gluon}
\end{equation}
where, $r_{q}$ is the radius of  quark,  $r_{s}$ and $r_{l}$ are respectively the radius and the length of the gluon tube. 

The density function under study here was taken to be~\cite{Kubiczek:2015zha},

\begin{equation} 
\begin{split}
\rho_{G-f} (\textbf{r} ; \textbf{r}_{1} ,\textbf{r}_{2} ,\textbf{r}_{3} ) & = N_{g} \frac{1 - \kappa}{3} \sum\limits_{ k = 1}^{3} \rho_{q}(\textbf{r} - \textbf{r}_{k} ; r_{q} ) + \\
                                                                                                     & N_{g} \frac{\kappa}{3} \sum\limits_{ k = 1}^{3} \rho_{g}[\mathcal{R}^{-1}[\theta_{k} ,  \phi_{k} ](\textbf{r} -     
                                                                                                          \frac{\textbf{r}_{k}}{2} ; r_{q} , \frac{r_{k}}{2}\textbf]
\end{split}
\label{rho_together}
\end{equation}   

where, $\mathcal{R[\theta , \phi ]}$ transforms vector (0,0,1) into $(\cos\phi \sin\theta , \sin\phi \cos\theta , \cos\theta) $ and $\textbf{r}_{k} = r_{k}(\cos\phi_{k} \sin\theta_{k} , \sin\phi_{k} \cos\theta_{k} , \cos\theta_{k}) $ (where, k = 1,2 and 3) is the position vector of $k^{th}$ effective quark. $N_{g}$ is a collision energy dependent normalization factor for the density function of proton and accounts for the number of partons inside a proton. One can obtain this number by confronting the estimations to experimental observables. The free parameter $\kappa$ allows to control the percentage of gluon body content and here it is taken to be 0.5  as a first approximation ~\cite{Kubiczek:2015zha}.This is the fraction of gluons (total number of gluons being $\kappa N_g$) out of all partons inside a proton at a given collision energy.

\subsection{Calculation of Thickness function and Overlap function}
\label{a}
The collision plane is taken to be in x-y hence dependence along z axis is integrated out as follows
\begin{equation}
T(x,y) = \int \rho(x,y,z) dz
\label{thickness_definition}
\end{equation}  
The calculated thickness function for the $\rho_{G-f}$ is 
\begin{equation}
\begin{aligned}
\begin{split}
T (x ,y) & = \sum\limits_{k=1}^{3} \frac{N_{g}}{3}  \frac{1 - \kappa}{2\pi r_{q}^2} e^{-l_{k}} + \frac{N_{g}\kappa}{3} (\frac{1}{(2\pi)^{3/2}r_{s}^2r_{l}} \\
            & \sqrt{\frac{\pi}{2}} (\frac{sin^2\theta_{k}}{2r_{s}^2} + \frac{cos^2\theta_{k}}{2r_{l}^2})^{-1/2})e^{-a_{k}(x-\frac{x_{k}}{2})^2} \\
            & e^{-b_{k}(y-\frac{y_{k}}{2})^2} e^{-c_{k}(x-\frac{x_{k}}{2})(y-\frac{y_{k}}{2})}
\end{split}
\end{aligned}
\label{thickness_calculated}
\end{equation} 

where, 
$r_{s} = r_{q}$ and $r_{l} = \frac{r_{k}}{2}$, for the present studies, we have taken $r_{q} = r_{s}$ = 0.25 fm following Ref. \cite{Kubiczek:2015zha},
\begin{equation}
l_{k} = \frac{(x-x_{k})^2 + (y - y_{k})^2}{2r_{q}^2}
\end{equation}
 and  
\begin{equation}
\begin{split}
a_{k} & = - cos^2\phi_{k}P_{k} + [\frac{1}{2r_{s}^2}(sin^2\phi_{k} + \\
         & cos^2\phi_{k}cos^2\theta_{k}) +\frac{1}{2r_{l}^2}(cos^2\phi_{k}sin^2\theta_{k})],
\end{split}         
\end{equation} 
\begin{equation}
\begin{split}
b_{k} &= - sin^2\phi_{k}P_{k} + [\frac{1}{2r_{s}^2}(cos^2\phi_{k} + sin^2\phi_{k}cos^2\theta_{k}) \\
         & +\frac{1}{2r_{l}^2}(sin^2\phi_{k} + sin^2\theta_{k})],
\end{split}
\end{equation} 
\begin{equation}
c_{k} = - sin^2\phi_{k}P_{k} [1 - 2[\frac{tan^2\theta_{k}}{r_{s}^2} + \frac{1}{r_{l}^2} ]]
\end{equation}
 and 
\begin{equation}
P_{k} = \frac{r_{l}^2 - r_{s}^2}{4(\frac{r_{l}^2}{cos^2\theta_{k}} + \frac{r_{s}^2}{sin^2\theta_{k}})}
\end{equation}
The overlap function $T_{pp}(b)$ for projectile proton (\textbf{A}) and target proton (\textbf{B}) is defined as
\begin{equation}
T_{pp}(b) = \int \int T_{A}(x - \frac{b}{2},y)T_{B}(x + \frac{b}{2},y) dxdy
\label{overlap_definition}
\end{equation}  
Here $T_{pp}$ is sum of 4-components namely quark-quark, quark-gluon, gluon-quark, gluon-gluon. Primed (unprimed) indices indicate variables corresponding to \textbf{B} (\textbf{A}). In the following, we provide overlap function for all the possible combination of partons.

\vspace{1em}

\textbf{A.1. The Quark-Quark term}

The overlap function for the interaction of two quarks:

\begin{equation}
\begin{split}
(T_{pp})_{qq}(b) &= \frac{N_{g}^2 (1-\kappa)^2}{36\pi r_{q}^2} \sum\limits_{k, k'=1}^{3}\\
                          & exp[-\frac{(b- x_{k} - x'_{k'})^2 - (y_{k} - y'_{k'})^2}{4r_{q}^2})]
\end{split}                          
\label{overlap_quark_quark}
\end{equation}

\textbf{A.2. The Gluon-Gluon term} 

The overlap function for the interaction of two gluon tubes:

\begin{equation}
(T_{pp})_{gg}(b) = \sum\limits_{k, k'=1}^{3} C_{k,k'}\sqrt{\frac{\pi}{\lambda_{k,k'}}}e^{-\frac{\gamma_{k,k'}^2}{4\lambda_{k,k'}}}
\label{overlap_gluon_gluon}
\end{equation} 

where,

\begin{equation}
\begin{split}
\gamma_{k,k'} & = \frac{c_{k} + c'_{k'}}{4(b_{k} + b'_{k'})}[c_{k}(b + x_{k}) - c'_{k'}(b-x_{k'}) \\
                       & + 2(b_{k}y_{k} - b'_{k'}y'_{k'})]
\end{split}
\label{overlap_quark_gluon_var1}
\end{equation}

\begin{equation}
\lambda_{k,k'}= (a_{k} + a'_{k'}) - \frac{c_{k} + c'_{k'}}{4(b_{k} + b'_{k'})}
\label{overlap_quark_gluon_var2}
\end{equation} 

\begin{equation}
\begin{split}
C_{k,k'} & = A_{k}A'_{k'}\sqrt{\frac{\pi}{b_{k} + b'_{k'}}}\\
             & exp[\frac{[\frac{1}{2}[c_{k}( b + x_{k} ) - c'_{k'}( b - x'_{k'})] + (b_{k}y_{k} + b'_{k'}y'_{k'})]^{2}} {4(b_{k} + b'_{k'})}]\\
             & exp[-\frac{a_{k}}{4}(b+x_{k})^2 - \frac{a'_{k'}}{4}(b- x'_{k'})^2 - \\
             & \frac{c_{k}y_{k}}{4}(b+x_{k}) + \frac{c'_{k'}y'_{k'}}{4}(b- x'_{k'})]
\end{split}
\label{overlap_quark_gluon_var3}
\end{equation}

\begin{equation}
%A_{k}= \frac{N_{g} \kappa}{3}\frac{1}{(2\pi)^{3/2}r_{s}^2r_{l}}(\frac{\pi}{2})^{1/2} [\frac{sin^2\theta_{k}}{2r_{s}^2} + \frac{cos^2\theta_{k}}{2r_{l}^2}]^{-1/2}
A_{k}= \frac{N_{g} \kappa}{3}\frac{1}{(2\pi)^{3/2}r_{s}^2r_{l}}(\frac{\pi}{2})^{1/2} [\frac{sin^2\theta_{k}}{2r_{s}^2} + \frac{cos^2\theta_{k}}{2r_{l}^2}]^{-1/2}
\label{overlap_quark_gluon_var3}
\end{equation} 

%\vspace{2em}

\textbf{A.3. The Quark-Gluon term} 

The overlap function for the interaction of a quark and a gluon tube:

\begin{equation}
(T_{pp})_{qg}(b) = \sum\limits_{k, k'=1}^{3} D_{k,k'}\sqrt{\frac{\pi}{\alpha_{k,k'}}}e^{\frac{\beta_{k,k'}^2}{4\alpha_{k,k'}}}
\label{overlap_quark_gluon}
\end{equation} 

where,

\begin{equation}
\alpha_{k,k'}= \frac{1}{2r_{q}^2} + a'_{k'} - \frac{(c'_{k'})^{2}}{4(\frac{1}{2r_{q}^2} + b'_{k'})}
\label{overlap_quark_gluon_var1}
\end{equation} 

\begin{equation}
\begin{split}
\beta_{k,k'} &= 2c'_{k'}[\frac{y_{k}}{r_{q}^2} + b'_{k'}y'_{k'} - \frac{c'_{k'}}{2}(b-x'_{k'}) \\
                  & - \frac{y'_{k'}}{4}] - \frac{\frac{b}{2} + x_{k}}{r_{q}^2} + a'_{k'}(b-x'_{k'})
\end{split}
\label{overlap_quark_gluon_var1}
\end{equation} 

\begin{equation}
\begin{split}
D_{k,k'} &= E_{k,k'}\sqrt{\frac{\pi}{\frac{1}{2r_{q}^2} + b'_{k'}}} \exp[-\frac{y_{k}^2}{2r_{q}^2} - \frac{b'_{k'}(y'_{k'})^2}{4}] \\
              & \exp[-\frac{(\frac{b}{2} + x_{k})^2}{2r_{q}  ^2} - \frac{a'_{k'}}{4}( b- x'_{k'})^2] \\ 
              & \exp[\frac{1}{4(\frac{1}{2r_{q}^2} + b'_{k'})}[\frac{y_{k}}{r_{q}^2} + b'_{k'}y'_{k'} \\
              & - \frac{c'_{k'}}{2}(b-x'_{k'})]^2]\exp[\frac{c'_{k'}y'_{k'}}{4}(b - x'_{k'})]
\end{split}
\label{overlap_quark_gluon_var3}
\end{equation}

\begin{equation}
E_{k,k'}= \frac{N_{g}^{2}\kappa(1 - \kappa)}{36{\pi}^{2}r_{q}^4 r_{k'}}[\frac{sin^2\theta'_{k'}}{2r_{q}^2} + \frac{2cos^2\theta'_{k'}}{r_{k'}^2}]^{-1/2}
\label{overlap_quark_gluon_var3}
\end{equation}

\vspace{2em}

\textbf{A.4. The Gluon-Quark term} 

The overlap function for the interaction of a gluon tube and a quark:

\begin{equation}
(T_{pp})_{gq}(b) = \sum\limits_{k, k'=1}^{3} F_{k,k'}\sqrt{\frac{\pi}{\delta_{k,k'}}}\exp[{\frac{\eta_{k,k'}^2}{4\delta_{k,k'}}}]
\label{overlap_Gluon_quark}
\end{equation} 

where,

\begin{equation}
\delta_{k,k'} = a_{k} + \frac{1}{2r_{q}^2} - \frac{c_{k}^2}{4(\frac{1}{2r_{q}^2} + b_{k})}
\label{overlap_quark_Gluon}
\end{equation} 

\begin{equation}
\begin{split}
\eta_{k,k'} &= a_{k}(b+ x_{k}) - \frac{1}{r_{q}^2}(\frac{b}{2} - x'_{k'}) - \frac{2c_{k}}{4(\frac{1}{2r_{q}^2} + b_{k})}\\
                & [b_{k}y_{k} + \frac{y'_{k'}}{r_{q}^2} +  \frac{c_{k}}{2}(b + x_{k})] 
\end{split}
\label{overlap_quark_Gluon}
\end{equation}

\begin{equation}
\begin{split}
F_{k,k'} & = G_{k,k'}\sqrt{\frac{\pi}{\frac{1}{2r_{q}^2} + b_{k}}} \exp[\frac{1}{4(\frac{1}{2r_{q}^2} + b_{k})}[b_{k}y_{k} + \\ 
            &  \frac{y'_{k'}}{r_{q}^2} + \frac{c_{k}}{2}(b + x_{k})]^2] \exp[ - \frac{c_{k}y_{k}}{4}(b + x_{k})- \\
            &\frac{{y'_{k'}}^2}{2r_{q}^{2}}-\frac{b_{k}y_{k}^{2}}{4}] \exp[- \frac{a_{k}}{4}(b + x_{k})^2 - \frac{1}{2r_{q}^2}(\frac{b}{2} - x'_{k'})^2]
%F_{k,k'}= G_{k,k'}\sqrt{\frac{\pi}{\frac{1}{2r_{q}^2} + b_{k}}} \exp[-\frac{1}{4(\frac{1}{2r_{q}^2} + b_{k})}[b_{k}y_{k} + \frac{y'_{k'}}{r_{q}^2} + \frac{c_{k}}{2}(b + x_{k})]^2] \exp[- \frac{b_{k}y_k^2}{4} - \frac{y'_k'^2}{2r_{q}^2} - \frac{c_{k}y_{k}}{4}(b + x_{k})] \exp[- \frac{a_{k}}{4}(b + x_{k})^2 - \frac{1}{2r_{q}^2}(\frac{b}{2} - x'_{k'})^2]
\end{split}
\label{overlap_quark_gluon_var3}
\end{equation}

\begin{equation}
G_{k,k'}= \frac{N_{g}^{2}\kappa(1 - \kappa)}{36{\pi}^{2}r_{q}^4 r_{k}}[\frac{sin^2\theta_{k}}{2r_{q}^2} + \frac{2cos^2\theta_{k}}{r_{k}^2}]^{-1/2}
\label{overlap_quark_gluon_var3}
\end{equation} 

Together total overlap function is sum of four terms given by Eq.~\ref{overlap_quark_quark}, \ref{overlap_gluon_gluon}, \ref{overlap_quark_gluon}, \ref{overlap_Gluon_quark}.

\begin{equation}
\begin{split}
T_{pp}(b) & = (T_{pp})_{qq}(b) + (T_{pp})_{gg}(b) + 
                (T_{pp})_{qg}(b) \\ & + (T_{pp})_{gq}(b)
\end{split}
\label{tpp}
\end{equation}

\subsection{Calculation of $N_{coll}$ and $N_{part}$}

We define the number of binary collisions ($N_{coll}$) of partons  in a $p+p$ collision at a given impact parameter (b) as follows: 
\begin{equation}
N_{coll}(b) = \sigma_{eff}T_{pp}(b),
\label{n_coll}
\end{equation} 
where $\sigma_{eff}$ is the effective partonic cross sections. It should be mentioned here that  quark-quark, quark-gluon and gluon-gluon interaction cross section will be different due to different color factors of quarks and gluons ~\cite{Combridge:1977dm}. However, we use a common partonic cross section here, which is extracted from the fits to the data as in Ref.~\cite{Drescher:2007cd}, which avoids limitations of theoretical calculations at the cost of loosing the information regarding difference in individual type of interaction. In the absence of experimental information and non-perturbative QCD based calculations of the individual cross-section ({\it e.g.} $gg, ~qg$ and $qq$ processes), we have taken a common cross-section for all partons as $\sigma_{eff}$.
In line with the previous studies~\cite{Kubiczek:2015zha,Glazek:2016vkl}, we fix $\sigma_{eff}$ = 4.3 $\pm$ 0.6 mb~\cite{Drescher:2007cd} with $N_{g}$= 10 partons, so as to reproduce the experimental value of inelastic cross section, $\sigma_{pp}$ = 60 mb~\cite{Chatrchyan:2012nj}  for $p+p$ collision at $\sqrt{s}$ = 7 TeV. This accounts for the only non-trivial dependence of the Glauber calculation on the beam energy $\sqrt{s}$. Previous studies~\cite{Drescher:2007cd,Kubiczek:2015zha} have assumed linear scaling of charged hadron ($N_{ch}$) multiplicity with $N_{coll}$ only. In contrast to this assumption, we have considered dependence of $N_{ch}$ on number of participants partons ($N_{part}$) and $N_{coll}$. Further, relationship between $N_{part}$ and $N_{coll}$ is considered non-linear as that of the heavy ion collisions assuming three dimensional shape. Thus, number of participating partons at impact parameter 'b' is given as
\begin{equation}
N_{part}(b) \propto N_{coll}^{1/x}(b)
\label{n_part}
\end{equation} 
where  x is a parameter.

By considering, $f$, as a fraction of charged hadron multiplicity produced from binary collisions, we have two component model for estimation of number of charged particles given as
\begin{equation}
\frac{dN_{ch}}{d\eta} = n_{pp}[(1 - f)\frac{N_{part}}{2} + f N_{coll}]
\label{dNchdEta}
\end{equation} 
where, $n_{pp}$ is a constant of proportionality, which represents the charged particle multiplicity density in pseudorapidity for $p+p$ collisions and $f$ is a free parameter. 

\section{Results}

Assuming initial position vectors of three quarks to be vertices of equilateral triangle in xy-plane as $\textbf{r}_{1} = (\frac{d}{4} , \frac{\sqrt{3}}{4}{d},0)$ , $\textbf{r}_{2} = (\frac{d}{4} , -\frac{\sqrt{3}}{4}{d},0) $ , $\textbf{r}_{3} = (-\frac{d}{2},0, 0)$, where, d is the free parameter of the model which ensures that the length of the gluon tubes connecting quarks are fixed i.e., ($|\textbf{r}_{1}|^{2}=|\textbf{r}_{2}|^{2}=|\textbf{r}_{3}|^{2}) = \frac{d^{2}}{4}$). For the present study, we have taken $d$ = 1.5 fm \cite{Kubiczek:2015zha}. Now, in order to account for all possible configurations, position vectors of quarks are parameterised by varying azimuthal and polar angles. Generalised configuration considering tilt by $\psi$ along x-axis and rotation by angle $\alpha$ are  $\textbf{r}_{1} = (\frac{d}{2}\cos(\frac{\pi}{3}+\psi) , \frac{d}{2}\sin({\frac{\pi}{3}+\psi}) \cos\alpha,-\frac{d}{2}\sin({\frac{\pi}{3}+\psi}) \sin\alpha)$,  $\textbf{r}_{2} = (\frac{d}{2}\cos(\frac{5\pi}{3}+\psi) , \frac{d}{2}\sin({\frac{5\pi}{3}+\psi}) \cos\alpha,-\frac{d}{2}(\sin{\frac{\pi}{3}+\psi}) \sin\alpha)$, $\textbf{r}_{3} = (\frac{d}{2}\cos(\psi) , \frac{d}{2}\sin{\psi} \cos\alpha,-\frac{d}{2}\sin{\psi} \sin\alpha)$ and considering tilt by $\gamma$ along y-axis and rotation by angle $\beta$ are $\textbf{r}_{1} = (\frac{d}{2}\cos(\frac{\pi}{3}+\gamma)\cos\beta, \frac{d}{2}\sin({\frac{\pi}{3}+\gamma}) ,\frac{d}{2}\cos({\frac{\pi}{3}+\gamma}) \sin\beta)$, $\textbf{r}_{2} = (\frac{d}{2}\cos(\frac{5\pi}{3}+\gamma)\cos\beta , \frac{d}{2}\sin({\frac{5\pi}{3}+\gamma}),-\frac{d}{2}\cos({\frac{\pi}{3}+\gamma}) \sin\beta)$, $\textbf{r}_{3} = (\frac{d}{2}\cos(\gamma)\cos\beta , \frac{d}{2}\sin{\gamma},\frac{d}{2}\cos{\gamma} \sin\beta)$.\\
 
In the above configurations, $\psi$ and  $\gamma$ $\epsilon$ (0,$\frac{2\pi}{3})$, $\alpha$ $\epsilon$ (0,$\pi$) and $\beta$ $\epsilon$ (0,2$\pi$).
In our present study, we have taken, x, in Eq.~\ref{n_part} to be 0.75 as $N_{coll}$  scales as A$^{4/3}$ for similar target and projectile nuclei with mass numbers A for heavy ion collisions and are spherical in shape. In our work, this consideration of x = 0.75 holds good because when the plane formed by connecting centres of each quark is randomly rotated as part of Monte Carlo simulation for accounting all possible configurations of collision geometry, the overall angular space is exhausted thus making collisions geometry to be closely spherical overlap with preserving contributions from each of the different configurations hence the factor 0.75 is taken so that it accounts for general spherical overlap in heavy-ion collisions. We have also chosen, RMS radius of proton and quark as 1 fm and 0.25 fm, respectively. 

\subsection{\textbf{Number of binary collisions and participants as a function of impact parameter}}

\begin{figure}
  \includegraphics[scale=0.401]{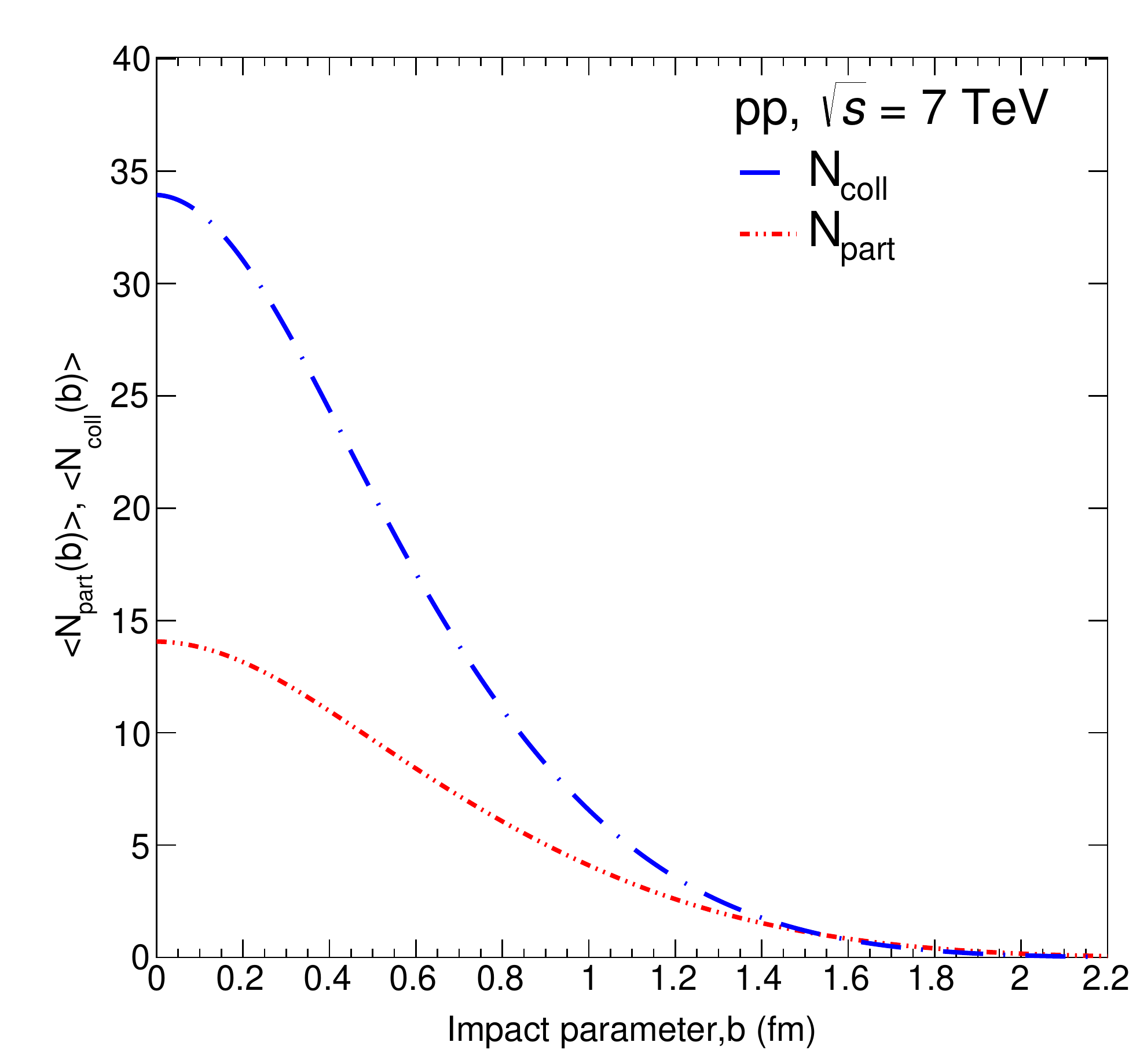}
\caption{(Color online) $N_{coll}$ and $N_{part}$ as a function of impact parameter (b) for different number of terms contraction}
\label{fig1}     
\end{figure}

We have used Eqs.~\ref{n_coll} and~\ref{n_part}, to estimate $N_{coll}$  and $N_{part}$. Fig.\ref{fig1} shows the mean value of $N_{coll}$ (upper curve) and $N_{part}$ (lower curve) as a function of impact parameter $(b)$. Towards higher values of $b$, the difference between the two curves effectively vanishes. Similar trends were observed for Au+Au and Cu+Cu collisions at $\sqrt{s_{NN}}$ = 200 TeV~\cite{Miller:2007ri}.

\subsection{\textbf{Charged particle multiplicity estimation}} 

Two-component models have been used in heavy-ion phenomenology for long time to estimate the charged-particle multiplicity~\cite{Wang:2000bf,Kharzeev:2000ph}. The inelastic cross section, $\sigma^{inel}_{NN}$, which depends on collision energy, is used as input for the MC Glauber model. In our current study, we have used similar approach for $p+p$ collisions as well, where nucleons are replaced by partons (quarks and gluons) and $\sigma^{inel}_{NN}$ by $\sigma_{eff}$. The model provides $N_{part}$ and $N_{coll}$, for an event with a given impact parameter and collision energy which is discussed in the previous section. 
%The particle multiplicity per parton-parton collision is parametrized by a Negative Binomial Distribution (NBD). 
As in heavy-ion collisions, the concept of ``ancestors" (independently emitting sources of particles) has been introduced for a given value of $N_{part}$ and $N_{coll}$. The number of ancestors can be parametrized by a two-component model given by~\cite{Wang:2000bf,Kharzeev:2000ph},
\begin{equation}
N_{ancestors} = f N_{part} + (1-f) N_{coll}
\label{ancestor}
\end{equation} 

\begin{figure}
  \includegraphics[scale=0.41]{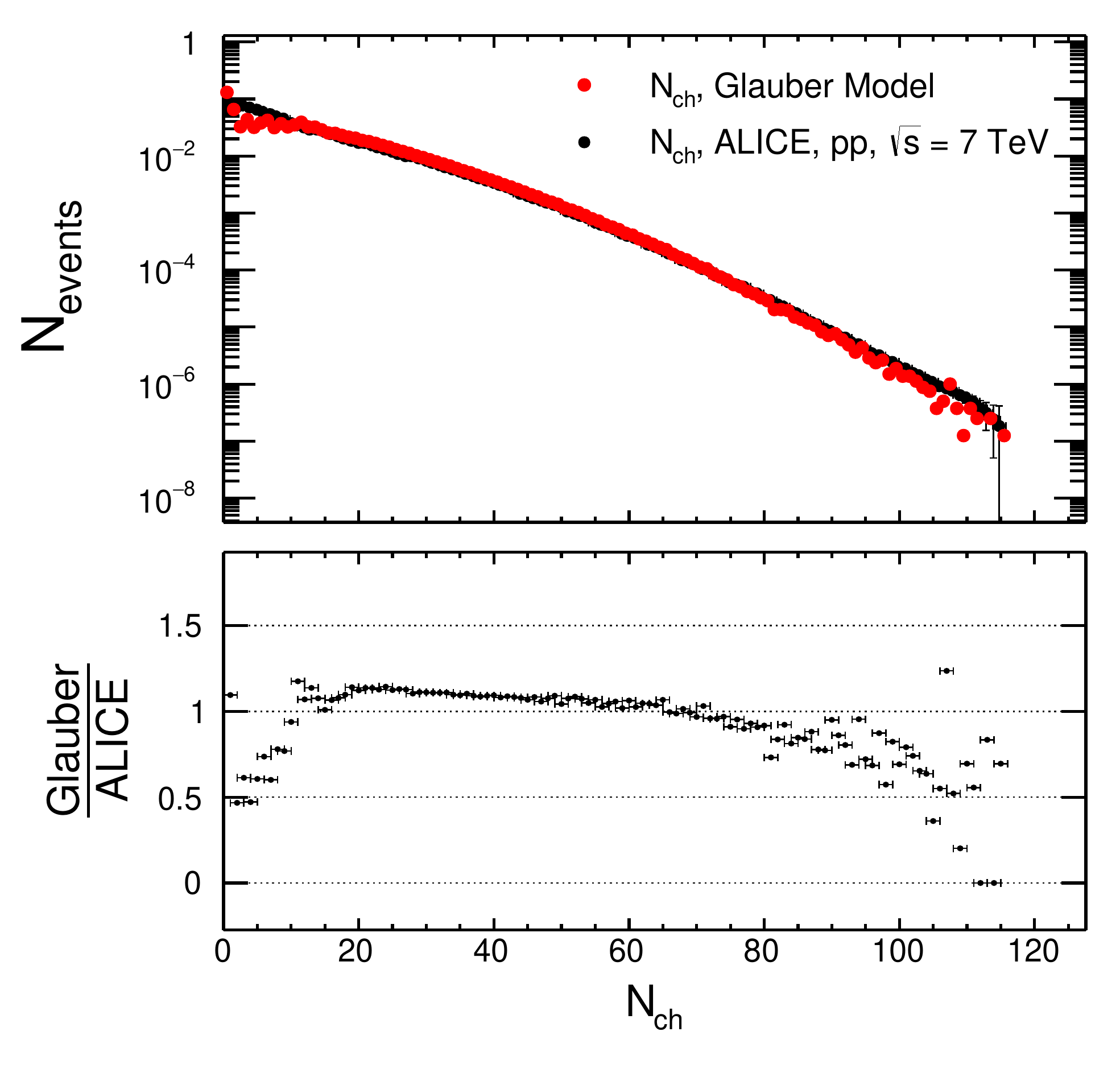}
  \caption{(Color online) Upper panel: Comparison of charged multiplicity distribution obtained from present work and ALICE experiment for $p+p$ collisions at $ \sqrt{s} =$ 7 TeV~\cite{Aamodt:2010pp}. Black dots represents ALICE data and red dots represent present work. Lower panel: Ratio of this work to the ALICE experimental data.}
   \label{fig:alicedata_glouber}     
  \end{figure}

The two-component model divides the parton-parton collisions into $soft$ and $hard$ interactions: the multiplicity of particles produced by soft interaction is proportional to $N_{part}$ and hard interaction is proportional to $N_{coll}$. As negative binomial distribution (NBD) able to well reproduce the charged-particle distribution in $p+p$ collisions~\cite{Alner:1985rj}, we use two-parameter NBD to calculate the probability of producing $n$ particles per ancestor;
  
\begin{equation}
P(n; \bar{n},k)=\frac{\Gamma(n+k)}{\Gamma(k)\Gamma(n+1)}
\left[\frac{\bar{n}}{k+\bar{n}}\right]^n
\left[\frac{k}{k+\bar{n}}\right]^k,
\label{eq:NBD}
\end{equation}
where $\bar{n}$ is the average multiplicity and $k$ characterizes the width of the distribution. By the use of different combination of $f$ (Eq.~\ref{ancestor}), $\bar{n}$ and $k$ ( Eq.~\ref{eq:NBD}) we have repeated the process of obtaining the multiplicity distribution for a large sample of events, until our model simulate the experimental multiplicity distribution. We have also calculated the ratio of $N_{ch}$ obtained from our model to that of experimental value and is represented in Fig.~\ref{fig:alicedata_glouber} for $p+p$ collisions at $ \sqrt{s} = 7$ TeV. The best agreement for $N_{ch}$ distribution obtained by our model with experimental data is found for $f=$ 0.85, $\bar{n} = 8$ and $k = 0.13$.  From Fig.~\ref{fig:alicedata_glouber}, it can be seen that our model well describes the data in the mid multiplicity region ( $15 < N_{ch} < 90$ ), with 5-10 \% discrepancy. However, towards the low and high multiplicity it is unable to reproduce the experimental measurement. The inability of the model to explain the extreme low and high multiplicity region might be due to the lower probability of collision impact parameter, when derived in the Monte Carlo from a Gaussian distribution.
%lack of statistics. 

\subsection{\textbf{Centrality estimation}} 

\begin{figure} [h]
  \includegraphics[scale=0.42]{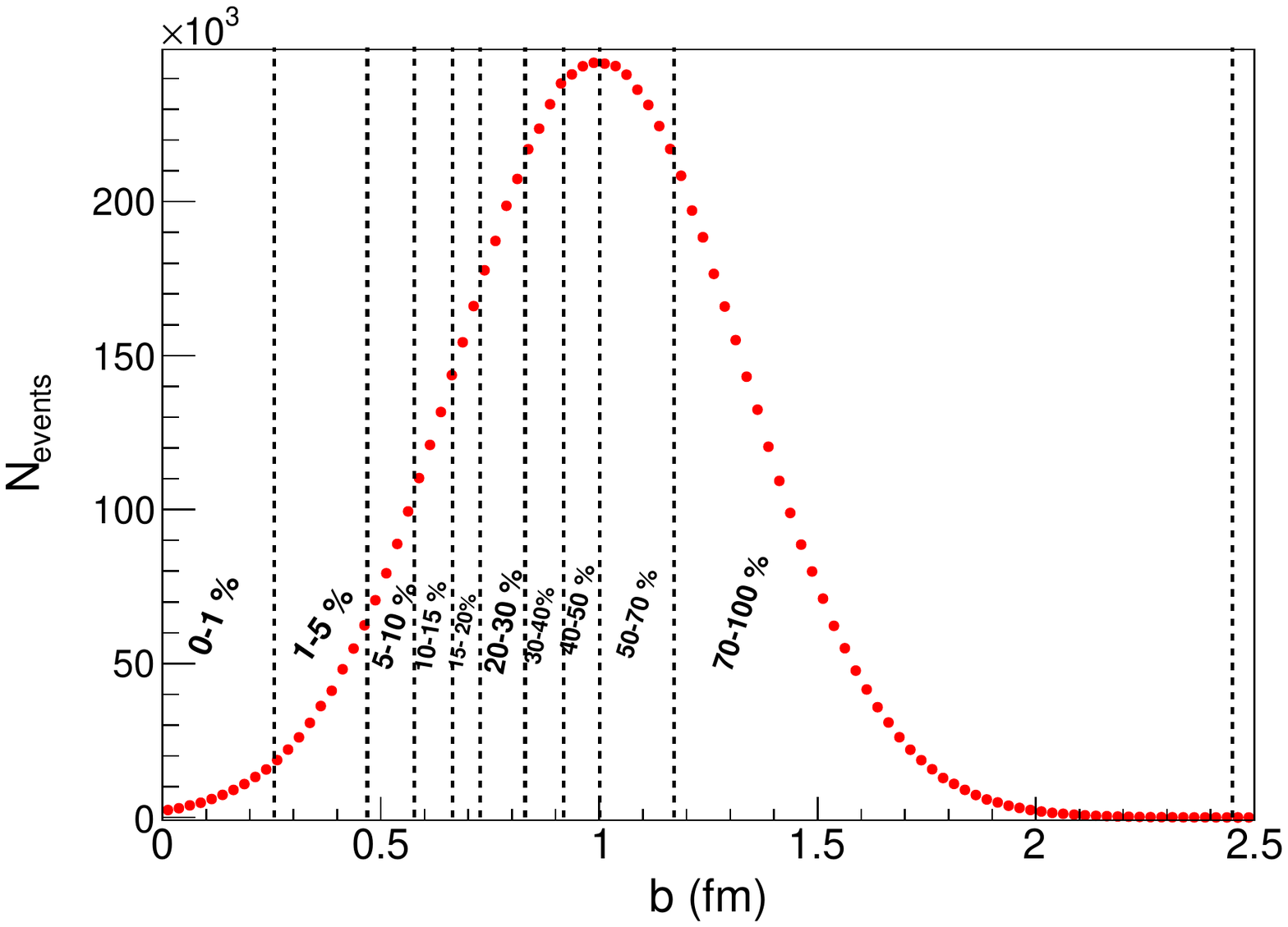}
  \caption{(Color online) Input impact parameter (b) profile for $p+p$ collisions.}
   \label{fig:b}     
  \end{figure}

The centrality is usually expressed as a percentage of the total interaction cross section, $\sigma$~\cite{Abelev:2013qoq}. Impact parameter distribution is taken as input to our current model. So, the centrality percentile of a $p+p$ collision with $b$ is defined by integrating the impact parameter distribution as,

\begin{equation}
c_{1} = \frac{\int_{0}^{b1} dN/db~db} {\int_{0}^{\infty} dN/db~db},~c_{2} = \frac{\int_{b1}^{b2} dN/db~db} {\int_{0}^{\infty} dN/db~db},.......
\label{percentile}
\end{equation} 
where $c_{1}$, $c_{2}$ ....etc.  are the percentile bins and $b_1, b_2$,...etc,. are the impact parameters. More clearly, $c_{1}$ percentage of total number of events of impact parameter distribution fall in the interval ($b_1, b_2$) and so on. For the current analysis, a Gaussian distribution with mean $1$ and standard deviation of $0.32$ has been used as input impact parameter distribution, which is shown in Fig.~\ref{fig:b}, so that the distribution function vanishes beyond the proton radius ($\approx$ 2 fm).  

\begin{figure} [h]
  \includegraphics[scale=0.44]{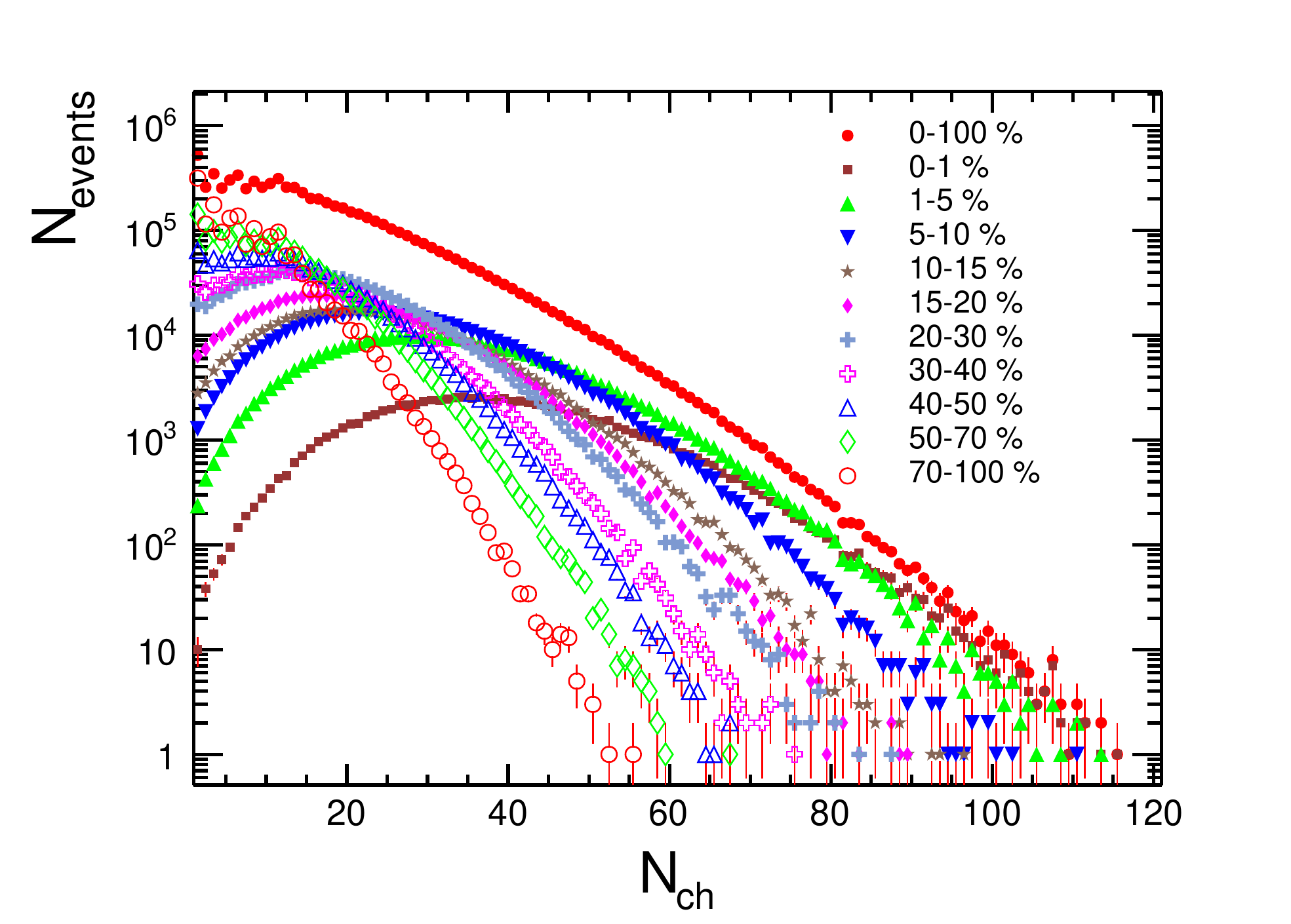}
  \caption{(Color online) Charged-particle multiplicity distribution in different percentile bins for $p+p$ collisions at $\sqrt{s}$ = 7 TeV.}
   \label{fig:centrality}     
  \end{figure}

We have also tested different forms of impact parameter distributions, but Gaussian distribution is found to be a suitable choice to describe the charged-particle multiplicity distribution. Once, we get the ranges of the impact parameter corresponding to each centrality, we have projected it to $N_{ch}$, $N_{part}$ and $N_{coll}$ to calculate $\langle N_{ch} \rangle$, $\langle N_{part} \rangle$ and $\langle N_{coll} \rangle$ corresponding to each b-ranges. Fig.~\ref{fig:centrality} represents the multiplicity distribution for each percentile bin. Table~\ref{table_final} shows the value of  $\langle N_{ch} \rangle$, $\langle N_{part} \rangle$ and $\langle N_{coll} \rangle$, obtained by using our model along with $\langle N_{ch} \rangle$ value of ALICE for pp collisions at $\sqrt{s} = 7$ TeV. 

  It can be clearly seen that the calculated $\langle dN_{ch}/d\eta \rangle$ is well consistent with experimental value, except for the high and low multiplicity regions. This is because of the artefact of incapability of our model to describe the charged-particle distribution in that region (Fig.~\ref{fig:alicedata_glouber}). However, it is to be noted that the input $\sigma^{inel}_{gg}$ = 0.43 $\pm$ 0.06 $fm^{2}$ contain 14 \% uncertainty and the same amount of uncertainty (14 \%) is associated with each $\langle dN_{ch}/d\eta \rangle$. From our model, we found  $\langle dN_{ch}/d\eta \rangle = 7.47$ for minimumbias (0-100 \%) collisions, which is little higher from the experimental value, $\langle dN_{ch}/d\eta \rangle =$ 6.01  $\pm$ $0.01^{+0.20}_{-0.12}$~\cite{Abelev:2012rz}. This discrepancy needs to be understood.

\begin{table*}[h]
\centering
\caption{Geometric properties ($b, N_{ch}, N_{part}, N_{Coll}$) of $p+p$ collisions for different multiplicity classes using Glauber Monte Carlo calculation along with a Negative binomial distribution fit to charged particle multiplicity distribution at $\sqrt{s}$ = 7 TeV for the ALICE experiment at the LHC.}
\small
\begin{center}
\begin{tabular}{|c|c|c|c|c|c|}
\hline
 Multiplicity (\%) &    $ b-range (fm)$      &  $\langle dN_{ch}/d\eta \rangle^{glauber}$  &  $\langle dN_{ch}/d\eta \rangle^{expt}$     &  $\langle N_{part}\rangle$     & $\langle N_{coll} \rangle$    \\ \hline
     0-1    	  &0 - 0.25534				  &19.69                       			         & $28.82^{+0.86}_{-0.84}$                  		&13.142 				 &31.156                                  \\\hline
  1-5    	&0.25535 - 0.46909                     & 16.24	               		                          & $20.34^{+0.58}_{-0.57}$      			        &11.164		 		  &24.815                                  \\ \hline
  5-10    	&0.46909 - 0.58484                     &13.37				                          & $15.80^{+0.34}_{-0.32}$	                         &9.244	   		           &19.478                                  \\ \hline
 10-15   	&0.58484 - 0.66430                     &11.61                             		         & $13.07^{+0.24}_{-0.22}$ 			        &8.037				   &16.153                                  \\ \hline
 15-20  	 &0.66431 - 0.72766                    &10.28	                           		        & $11.25^{+0.19}_{-0.18}$ 		                &7.131    			           &13.818	                                  \\ \hline
 20- 30 	 &0.72767 - 0.83026                    &8.94                            	                        & $9.21^{+0.15}_{-0.14}$   				        &6.116   			            &11.326                                  \\ \hline
 30-40  	&0.83027 - 0.91819                  	  &7.48                                                       & $7.13^{+0.12}_{-0.11}$		        	        	        &5.268			            &9.215                                     \\ \hline
 40-50       &0.91820 - 1.00117   		  &6.49						       & $5.65^{+0.11}_{-0.09}$					&4.418				    &7.340                                     \\ \hline
 50-70       &1.00118 - 1.17163			  &5.12						        & $3.81^{+0.07}_{-0.06}$				        &3.395			             &5.208                                     \\ \hline
 70-100     &1.17164 - 2.54998			  &3.66						         & $1.66^{+0.05}_{-0.04}$		                         &1.968			             &2.591                                      \\ \hline
 
\end{tabular}
\end{center}
\label{table_final}
\end{table*}

  \vspace{0.5cm}

 %\vspace{0.02em}
\subsection{\textbf{The ratio, $R_{HL}$ for high to low multiplicity events}}
\begin{figure}
  \includegraphics[scale=0.405]{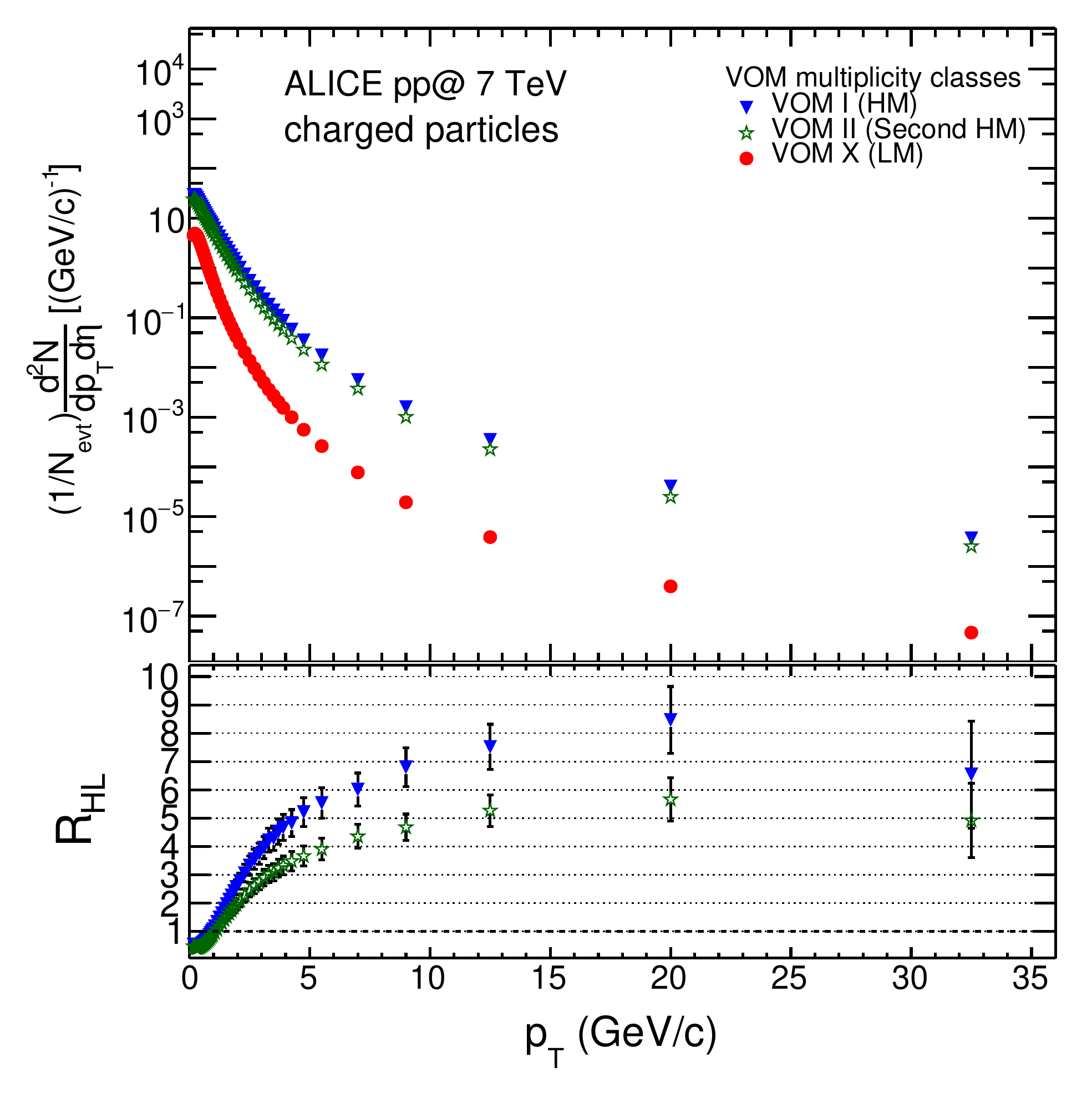}
  \caption{(Color online) Upper panel: Transverse momentum spectra of charged particle in $p+p$ collisional at $\sqrt{s}$ = 7  TeV~\cite{Acharya:2018orn} for VOM multiplicity classes, viz., highest (HM), second highest (second HM) and lowest multiplicity (LM) class. Lower Panel: $R_{HL}$  obtained from the ratio of differential yield at high-multiplicity and second high multiplicity classes with low multiplicity class scaled by $\langle N_{coll} \rangle$.  }  
 \label{fRpp}     
  \end{figure}

In order to understand the possibility of formation of a medium in high-multiplicity events in $p+p$ collisions, we define a variable as:

\begin{equation}
 R_{HL}(p_{T}) = \frac{d^{2}N/dp_{T}d\eta|^{HM}}{d^{2}N/dp_{T}d\eta|^{LM}} \times \frac{\langle N_{coll}^{LM} \rangle}{\langle N_{coll}^{HM} \rangle}
\label{Rpp}
\end{equation} 

which is similar to the nuclear modification factor $R_{AA}$ in heavy-ion collisions. Here, $d^{2}N/d\eta dp_{T}|^{HM}$, $d^{2}N/d\eta dp_{T}|^{LM}$, $\langle N_{coll}^{LM} \rangle$ ($\langle N_{coll}^{HM} \rangle$) are charged particle yields in high-multiplicity, low-multiplicity $p+p$ collisions at $\sqrt{s}$ = 7  TeV~\cite{Acharya:2018orn}, mean number of binary collisions in low (high) multiplicity $p+p$ events respectively. Upper panel of Fig.~\ref{fRpp} shows the transverse momentum spectra of charged particle in high-multiplicity (VOM I), second high multiplicity (VOM II) and low multiplicity (VOM X) events obtained from Ref.~\cite{Acharya:2018orn}. And lower panel shows the $R_{HL}$ defined in Eq.~\ref{Rpp}. For such definition of $R_{HL}$, it is observed for all charged particles for $p_{T} <$ 1 GeV/c, value of $R_{HL} <$ 1 and for $p_{T} >$ 1 GeV/c , it is greater than 1. However, it tends to reduce at very high $p_{T}$. And for $p_{T}>$1 GeV, the value of the factor is higher for higher multiplicities.

\begin{figure}
  \includegraphics[scale=0.41]{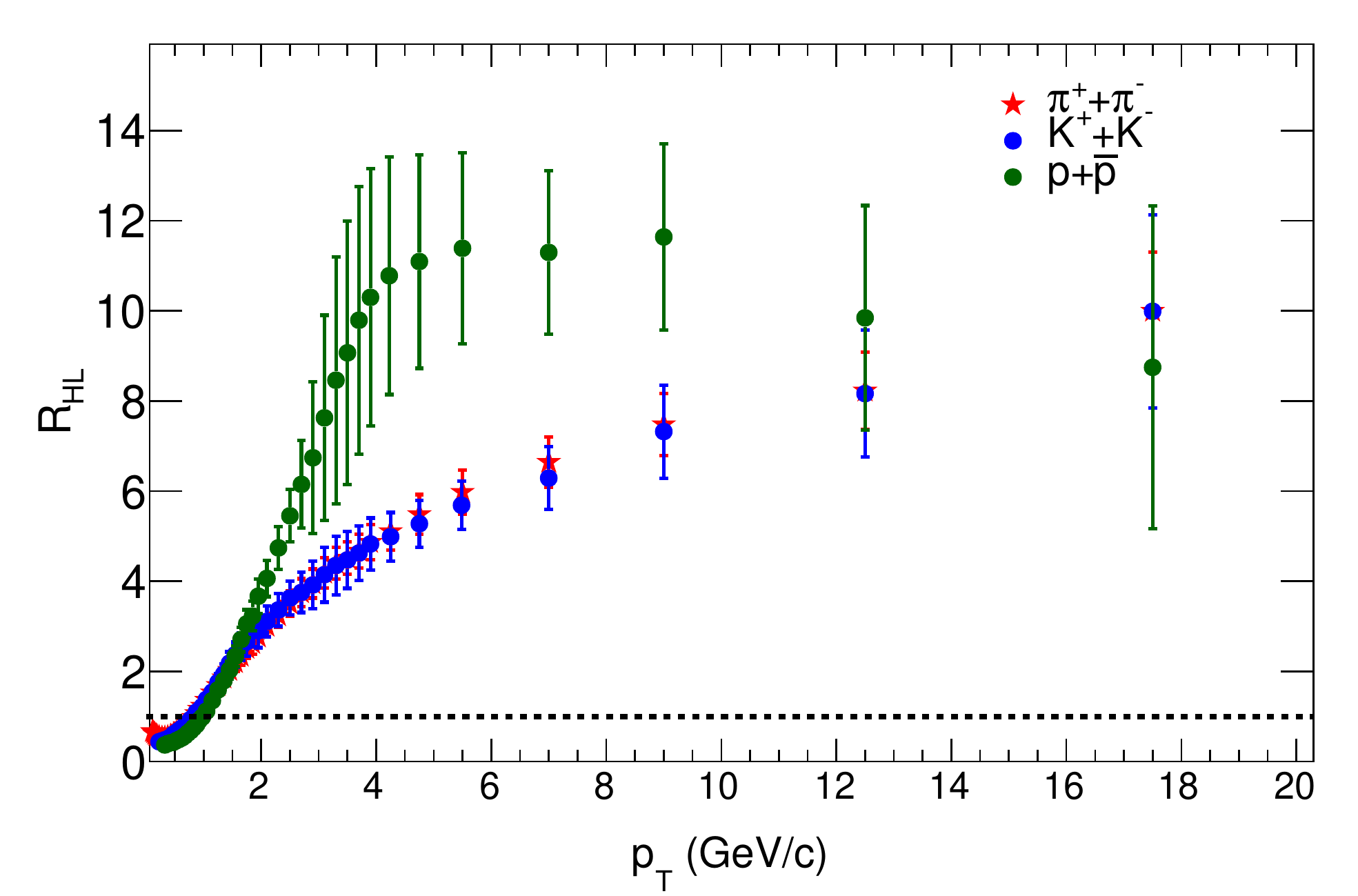}
  \caption{(Color online) Nuclear modification-like factor obtained from Eq.~\ref{Rpp} for pion, Kaon and proton in $p+p$ collisional at $\sqrt{s}$ = 7  TeV~\cite{Acharya:2018orn}.}
   \label{fRpp_pion_kaon_proton}     
  \end{figure}

Fig.~\ref{fRpp_pion_kaon_proton} shows results of $R_{HL}$ for identified particles, pion ($\pi^{+}+\pi^{-}$), Kaon ($K^{+}+K^{-}$), proton ($p+\bar{p}$) for $p+p$ collisions at $\sqrt{s}$ = 7. It is found that $R_{HL} < 1$ for proton for $p_{T} <1$ GeV which is same as observed in case of charged particles. However for pion and kaon $R_{HL} < 1$ for $p_{T} <0.8$ GeV. It is also observed that for $p_{T}< 1.9$ GeV, this identified particles have same almost value of $R_{HL}$  and for $p_{T}>1.9$ GeV, the value is almost same for pion and kaon but the value for the proton is larger and increases with $p_{T}$ sharply upto $p_{T} = 5$ GeV, and then saturates within uncertainties. But for pion and kaon, the factor increases monotonically with decreasing slope from $p_{T}>1.9$ GeV,  where the trend splits for proton and other two hadrons. 

It is reported that~\cite{Bencedi:2016tks}, proton shows distinct behaviour in this regard than other hadrons produced in p-Pb collisions. Also for p-Pb collisions, it is reported that the factor, $R_{pPb}>$ 1,  for all charged particle for $p_{T} >$ 2.5 GeV~\cite{ Bencedi:2016tks,Khachatryan:2016odn}. For p-Pb, $R_{pPb}$ saturates to unity for $p_{T} >$ 2 GeV, and it is also found that for $p+p$, $R_{HL}$ shows almost similar trend but with larger value of the factor with saturation-like behaviour starting after $p_{T} =$ 2 GeV. We note that the $R_{HL}$ values above unity for $p_T >$ 1 GeV may be qualitatively similar to other observed enhancements due to the Cronin effect and radial flow in $pA$ and $dA$ systems~\cite{Adare:2013esx,Khachatryan:2015waa}, as conjectured for similar behaviour of $R_{pPb} $~\cite{Khachatryan:2016odn}, where the moderate excess at high $p_{T}$ is suggestive of anti-shadowing effects in the nuclear parton distribution function~\cite{Arneodo:1992wf}.
 
\subsection{\textbf{Estimation of Elliptic-flow}}

\begin{figure}
  \includegraphics[scale=0.403]{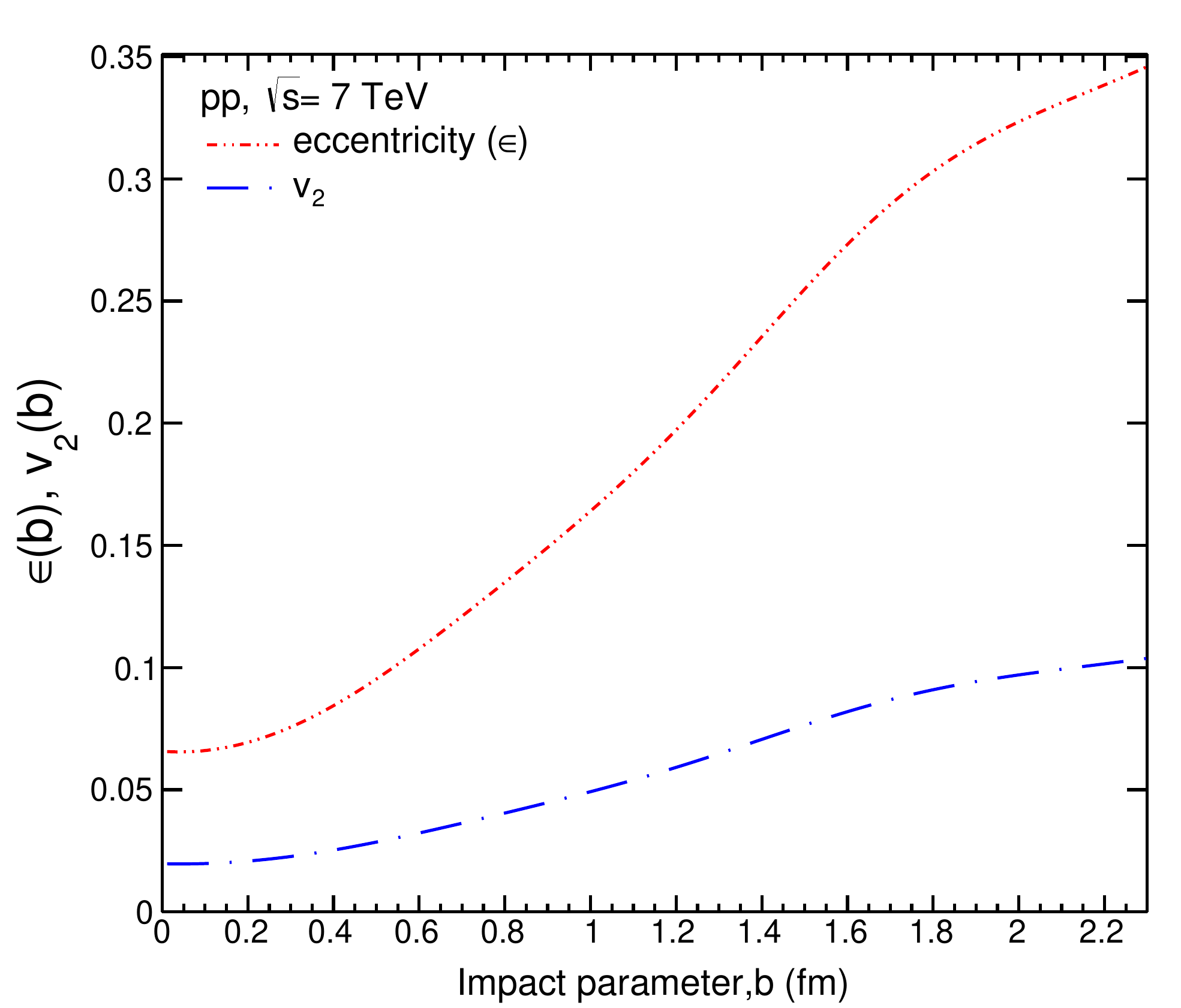}
  \caption{(Color online) Eccentricity ($\epsilon$) and elliptic-flow ($v_{2}$) as a function of impact parameter in $p+p$ collisions at $\sqrt{s}$ = 7  TeV.}
   \label{fig4}     
  \end{figure}

For long time, $p+p$ collisions were considered as the baseline measurements for determination of deconfined state of matter i.e., QGP formed in nuclear collision. Recent observation of $p+p$ collisions at LHC energies hints toward collective effect, thus, it becomes imperative to review the earlier view. In this regard, we have also calculated eccentricity ($\epsilon$) using the present approach. Asymmetry ratio between semi-axis dimensions of the overlap region weighted by $N_{coll}$ at a particular $b$ can be used to obtain $\epsilon$ as ~\cite{Glazek:2016vkl}:

\begin{equation}
 \epsilon(b) = \frac{\int  (y^2 - x^2)n_{coll}(x,y,b)dxdy}{\int  (y^2 + x^2)n_{coll}(x,y,b)dxdy}
\label{eccentricity}
\end{equation} 
where, $n_{coll}(x,y,b) = \sigma_{gg} T_{a}(x - \frac{b}{2},y)T_{b}(x + \frac{b}{2},y)$ represents impact plane binary collision density. We have calculated $\epsilon(b)$  by using Eq.~\ref{eccentricity} by considering sum of 4-components namely quark-quark, quark-gluon, gluon-quark and gluon-gluon. Fig.~\ref{fig4} shows the eccentricity for $p+p$ collision at $\sqrt{s}$ = 7 TeV obtained using Eq.~\ref{eccentricity} and it is observed to increase with $b$  and seems to saturate towards larger $b$.

Using $\epsilon$, we have obtained elliptic flow ($v_{2}$) as a function of $b$ by considering the scaling : $v_{2} = \Omega \epsilon$, where $\Omega$ = 0.3 $\pm$ 0.02~\cite{Drescher:2007cd}. Although we have considered a linear scaling to understand the variation of v2 with multiplicity, as a matter of fact v2 should be calculated by using relativistic hydrodynamics with relevant initial conditions and equation of state as inputs. 

 By geometry,  $v_{2}(b)$ will follow the general trend of $\epsilon$(b). It is found that the overlap of two hard spheres with infinitely
sharp edges yields artificially large eccentricities ~\cite{Voloshin:2008dg}.

\begin{figure}
  \includegraphics[scale=0.41]{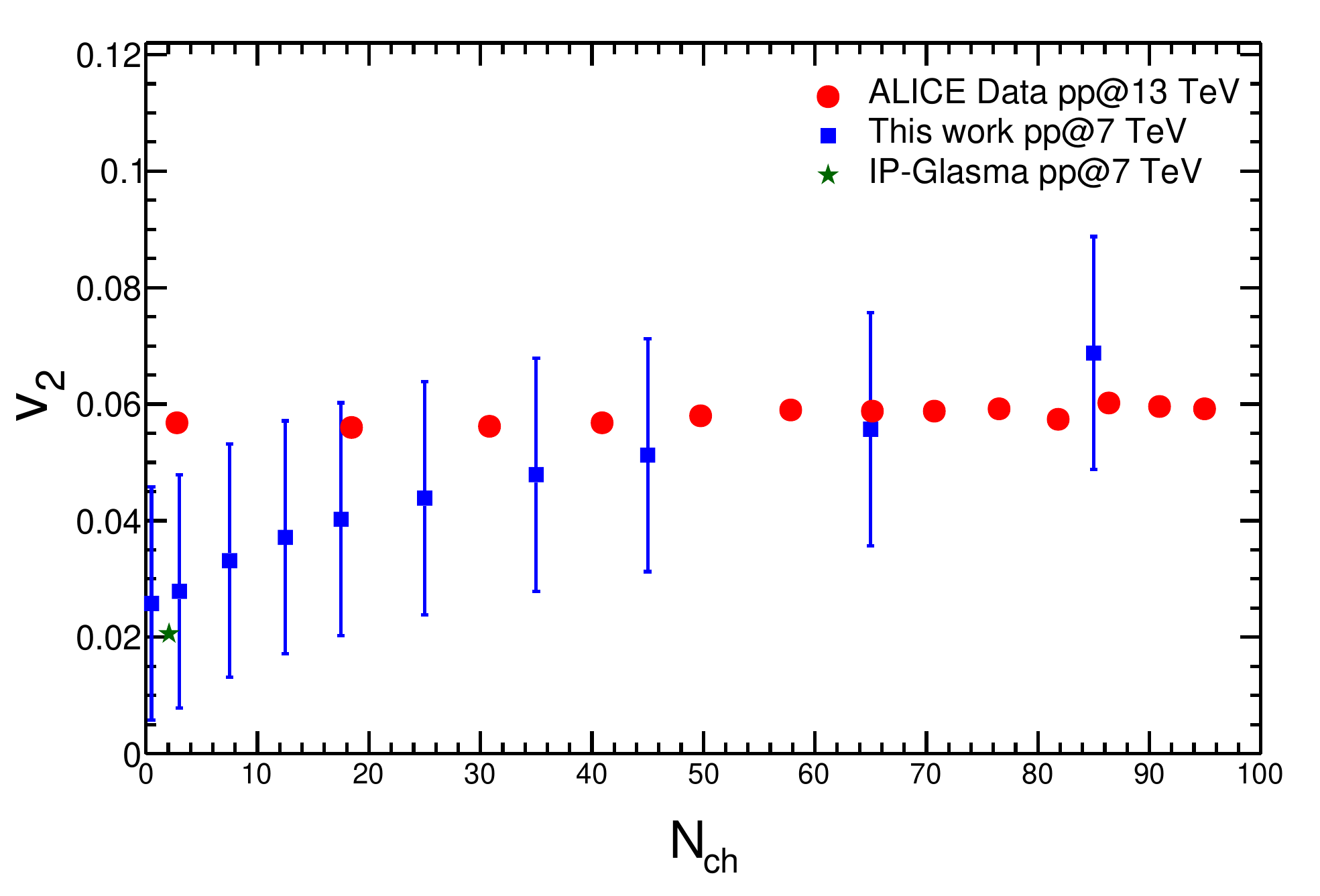}
  \caption{(Color online) Elliptic-flow, $v_{2}$  as a function of multiplicity in $p+p$ collisions at LHC energies.}
   \label{fig6}     
  \end{figure}

In Fig.~\ref{fig6}, we have compared our estimation of variation of $v_{2}$ with charged particle multiplicity for $p+p$ collision at $\sqrt{s} =$7 TeV with experimental result at $\sqrt{s} =$13 TeV~\cite{Gajdosova:2018pqo}. This is due to the fact that, the data for collisions at $\sqrt{s} =$13 TeV was not available at the time of reporting of this work to constrain our model. That does not prevent us from the comparison, since in Ref.~\cite{Aad:2015gqa}, it is reported that value of $v_{2}$ for collisions at $\sqrt{s} =$2.76 TeV and $\sqrt{s} =$13 TeV are almost the same when measured for different transverse momentum, indicating that the collision energy dependence of $v_{2}$ is weak. It is observed that for $N_{ch} \gtrsim$ 8, our estimation of $v_{2}$ with linear response to initial geometry reproduces the value obtained from experiment within the error bars. However, for lower multiplicities, our estimation with linear response to initial eccentricity falls short to that obtained from experimental data. This may be due to effects other than collective linear response or final state effects. Though, the charged particle multiplicity variation of $v_2$ for $p+p$ collisions at $\sqrt{s} =$ 7 TeV is not available, the elliptic flow coefficient extracted from the CMS Collaboration data at $\sqrt{s} =$7 TeV is $0.04-0.08$~\cite{Bozek:2010pb} and our estimation of $v_2$ falls within this range. We also note that this model gives $v_{2}$ similar  to that of the IP-Glasma model as presented in Ref~\cite{Bzdak:2013zma} for low multiplicity region ($< 8$).

\section{Summary and Discussion}

%What is new
In this work, we have investigated predictions of Glauber model for the initial condition for $p+p$ collisions, which considers anisotropic and inhomogeneous proton density profile. The results have been contrasted with experimental data. This model for density profile is inspired by the structure function obtained from deep inelastic scattering. Instead of distributing the positions of valence quarks randomly by keeping center of mass intact, we have taken random orientations generated by random rotation around three spatial axes, where center of  three quarks form a plane and connecting gluon tubes always remain fixed in length. This prevents the overlap of two valence quarks in space and possible placement of a quark out of proton radius,  where these two can happen for the first kind of randomization with only center of mass being fixed~\cite{Kubiczek:2015zha}, and to condition which may bring extra complication in the randomization process for not allowing it, generating ''spooky'' correlations. However, the present approach, apart from avoiding such complications, will give better handle for future investigations.
% on the role of the orientation of spin of the proton on result and estimations of the magnetic field produced in p+p collision (work in progress), which is expected to be present in collisions. 

With all these considerations, we have studied multiplicity distribution, to obtain the impact parameter to multiplicity relation, multiplicity dependence of initial eccentricity and azimuthal flow harmonics ($v_2$). It is found that this model can  well reproduce multiplicity distribution produced in $p+p$ events at ALICE, with the free parameter $f$ = 0.85. With properly constraining our model with experimental data and calibrating the range of $b$ with multiplicity percentile, we have used the estimated $\langle N_{coll} \rangle$ to obtain nuclear modification-like factor ($R_{HL}$) for $p+p$ collisions. It is found that the defined factor $< 1$ for $p_T <$ 1 GeV, and beyond this, the factor $> 1$. Moreover, it tends to reduce at very high $p_{T}$, and for $p_{T}>$ 1 GeV, the value of the factor is higher for higher multiplicities. We have also studied $R_{HL}$ for identified particles for $p+p$ collisions at $\sqrt{s}$ = 7 TeV, and found that the trend for $R_{HL}$ is similar to that observed in p-Pb system but with increased value. This behaviour at higher $p_{T}$ may be due to non-collective flow effects, which needs further investigation.

The non-availability of results from experiments which shows the variation of eccentricity and  $v_2$  with multiplicity at $\sqrt{s} =$ 7 TeV prevents us from comparing our estimation with experimental data at $\sqrt{s} =$ 7 TeV. However, we have compared our result of $v_2$ with that of $p+p$ collisions at $\sqrt{s} =$ 13 TeV, as it is observed that the collision energy dependence of $v_2$  is weak. We found that the result of $v_2$ obtained from present approach is in agreement with the result obtained in IP-Glasma model in lower multiplicity region. Also, it is found that the values of $v_2$ obtained from present model for $N_{ch} \gtrsim 8$ are very close to that of experimental data for $\sqrt{s} =$ 13 TeV. 

 The elliptic flow, $v_2$ measured through the anisotropic momentum distribution of produced 
particle is generated by the hydrodynamic pressure gradient resulted from the 
spatial anisotropy of the system formed initially. Therefore,  $v_2$ 
can be used to characterize the evolving medium and to do that, any momentum dependence resulting from other sources has to be subtracted out. The initial conditions required to solve the hydrodynamic equations are quantities 
which depend on the spatial coordinate but are momentum independent. Therefore,  
the initial condition obtained in the present study will be relevant for studying the evolving matter formed in $p+p$ collisions. 
The momentum dependent initial condition obtained in IP-Glasma model (e.g. the work reported in Ref. ~\cite{Schenke:2019pmk}) can also be useful to study hydrodynamic evolution 
when the  momentum dependence is integrated out.

\section{Acknowledgement} 
SD, GS and RNS acknowledge the financial supports  from  ALICE  Project  No. SR/MF/PS-01/2014-IITI(G) of Department  of  Science $\&$ Technology,  Government of India. DT acknowledges UGC, New Delhi, Government of India for financial supports. RNS acknowledges the financial supports from DAE-BRNS Project No. 58/14/29/2019-BRNS. Also, SD, GS and PS gratefully acknowledge Patryk Kubiczek for discussion on geometrical models for $p+p$ collisions.

\vspace{0.05cm}

\section*{Appendix: MULTIPLICITY VS GEOMETRIC
PROPERTIES OF THE COLLISION}
Table~\ref{table_final} shows geometric properties ($\langle b \rangle, \langle N_{ch} \rangle, \langle N_{part} \rangle, \langle N_{Coll} \rangle$) of $p+p$ collisions for different multiplicity classes using Glauber Monte Carlo calculation along with a Negative binomial distribution fit to charged particle multiplicity distribution at $\sqrt{s}$ = 7 TeV for the ALICE experiment at the LHC.

\end{document}